\documentclass[proof]{WileyASNA-v1}
\usepackage{moreverb}

\newcommand\BibTeX{{\rmfamily B\kern-.05em \textsc{i\kern-.025em b}\kern-.08em
T\kern-.1667em\lower.7ex\hbox{E}\kern-.125emX}}

\articletype{Article Type}%

\received{6 October, 2021}
\revised{6 October, 2021}
\accepted{7 October, 2021}

\begin{document}

\title{The gamma-ray emission from young radio galaxies and quasars}

\author[1,2,3]{Giacomo Principe*}

\author[4,5]{Leonardo Di Venere}

\author[3]{Giulia Migliori}
\author[3]{Monica Orienti}
\author[3]{Filippo D'Ammando}
\author[]{on behalf of the \textit{Fermi}-LAT collaboration}

\authormark{Giacomo Principe \textsc{et al}}

\address[1]{\orgdiv{Universit\'a di Trieste, Dipartimento di Fisica}, \orgaddress{\state{I-34127 Trieste}, \country{Italy}}}

\address[2]{\orgdiv{Istituto Nazionale di Fisica Nucleare, Sezione di Trieste}, \orgaddress{\state{I-34127 Trieste}, \country{Italy}}}

\address[3]{\orgdiv{Istituto Nazionale di Astrofisica - Istituto di Radioastronomia},  \orgaddress{\state{I-40129 Bologna}, \country{Italy}}}

\address[4]{\orgdiv{Dipartimento di Fisica “M. Merlin” dell’Università e del Politecnico di Bari}, \orgaddress{\state{I-70126 Bari}, \country{Italy}}}

\address[5]{\orgdiv{Istituto Nazionale di Fisica Nucleare, Sezione di Bari}, \orgaddress{\state{I-70126 Bari}, \country{Italy}}}

\corres{*Giacomo Principe, \email{giacomo.principe@ts.infn.it}}


\abstract[Abstract]{According to radiative models, radio galaxies are predicted to produce $\gamma$-rays from the earliest stages of their evolution onwards. The study of the high-energy emission from young radio sources is crucial for providing information on the most energetic processes associated with these sources, the actual region responsible for this emission, as well as the structure of the newly born radio jets. 
Despite systematic searches for young radio sources at $\gamma$-ray energies, only a handful of detections have been reported so far. Taking advantage of more than 11 years of \textit{Fermi}-LAT data, we investigate the $\gamma$-ray emission of 162 young radio sources (103 galaxies and 59 quasars), the largest sample of young radio sources used so far for a $\gamma$-ray study. We analysed the \textit{Fermi}-LAT data of each source separately to search for a significant detection. In addition, we performed the first stacking analysis of this class of sources in order to investigate the $\gamma$-ray emission of the young radio sources that are undetected at high energies. In this note we present the results of our study and we discuss their implications for the predictions of $\gamma$-ray emission from this class of sources.
 }


\jnlcitation{\cname{%
\author{Principe G..}, 
\author{L. Di Venere}, 
\author{G. Migliori}, 
\author{M. Orienti}, and 
\author{F. D'Ammando}} (\cyear{2021}), 
\ctitle{Gamma-ray emission from young radio galaxies and quasars}, \cjournal{AN}, \cvol{2021; 12:XX--XX}.}

\maketitle

\section{Introduction}\label{sec1}
One of the greatest questions investigated by modern astrophysics is understanding the origin of the $\gamma$-ray emission in radio galaxies and quasars.
While the extragalactic $\gamma$-ray sky is dominated by blazars, due to their small jet inclination to the line of sight and by relativistic beaming, only a few radio galaxies have been detected so far \citep[4LAC,][]{2020ApJ...892..105A}. Due to their misaligned jets, they offer a unique tool to investigate some of the non-thermal processes at work in unbeamed regions in active galactic nuclei (AGN).

According to the evolutionary scenario, the size of a radio galaxy is strictly related to its age \citep{1995A&A...302..317F}. Therefore extragalactic compact radio objects (i.e. radio objects with projected linear size LS<20 kpc), 
are expected to be the progenitors of extended radio galaxies \citep{1996ApJ...460..612R}.
Support to the young nature of these objects is given by the determination of kinematic and radiative ages in some of the most compact sources ($t \sim 10^2 - 10^5$ years) \citep{1982A&A...106...21P}. Compact radio objects reside either in galaxies or quasars. While for the latter the $\gamma$-ray emission is favored by their smaller jet-inclination angle and beaming effects, the origin of $\gamma$-ray emission in galaxies is still a matter of debate.

In the model proposed by \citet{2008ApJ...680..911S}, young radio galaxies are expected to produce isotropic high-energy emission through IC scattering of the UV photons by the electrons in the compact radio lobes. Depending on the sources' physical parameters (e.g. jet power, UV luminosity), the model predicted that young radio sources could constitute a class of $\gamma$-ray emitters detectable by \textit{Fermi}-LAT.
However, systematic searches for $\gamma$ rays from young radio sources have so far been unsuccessful \citep{2016AN....337...59D}.
Dedicated studies reported a handful of detections: three young radio galaxies (NGC\,6328 \citep{2016ApJ...821L..31M}, NGC\,3894 \citep{2020A&A...635A.185P} and TXS\,0128+554 \citep{2020ApJ...899..141L}),
and five compact steep spectrum (CSS) sources (3C\,138, 3C\,216, 3C\,286, 3C\,380, and 3C\,309.1) all associated with quasars \citep[4FGL,][]{2020ApJS..247...33A}.
The search for high-energy emission from young radio galaxies and quasars is crucial for investigating the energetic processes in the central region of the host galaxy, as well as the origin and the structure of the newly born radio jets.

Taking advantage of the increased sensitivity provided by more than eleven years of LAT data,
 we investigate the $\gamma$-ray properties of a sample of 162 young radio sources (103 galaxies and 59 quasars). In addition to the $\gamma$-ray analysis of each young radio object, we perform the first stacking analysis of this class of sources in order to investigate the $\gamma$-ray emission of the young radio sources still below the detection threshold in the high-energy regime.

Throughout this work, we assume $H_{0} = 70$ km s$^{-1}$ Mpc$^{-1}$ , $\Omega_{M} = 0.3$, and $\Omega_{\Lambda}=0.7$ in a flat Universe.

\section{Sample of young radio sources and analysis description} 
We selected the young radio sources for our study from the following resources which contain radio galaxies and quasars with LS<50 kpc: \citet{2009A&A...498..641D,2014MNRAS.438..463O, 2020MNRAS.491...92L, 2020ApJ...892..116W}. 
We also added to our sample the sources NGC\,3894, TXS\,0128+554, and 3C\,380, since they were already detected at high energy and investigated in \citet{2020A&A...635A.185P}, \citet{2020ApJ...899..141L} and \citet{2020ApJ...899....2Z}, respectively. Our final sample consists of 162 sources (103 galaxies and 59 quasars) with known position, redshift, LS, radio luminosity and peak frequency.

The selected sources have redshift values between 0.001 and 3.5 and LS spanning from less than 1 pc up to a few tens of kpc.
Most (129) of the sources have redshift below 1, with seven sources located in the local Universe ($z$ < 0.05, $D_\mathrm{L} \lesssim$200 Mpc). Considering the morphological classification, about half (79) of the sources are classified as compact symmetric objects (CSOs, LS $<1$ kpc), 70 sources as medium symmetric objects (MSOs, LS $\sim 1-20$ kpc), and 13 large symmetric objects (LSOs) with LS between 20 and 50 kpc.
Concerning their radio spectra and peak frequency ($\nu_\mathrm{p}$), 52 sources are classified as GHz-peaked spectrum (GPS,$\nu_\mathrm{p} > 0.5$ GHz), with the remaining 110 being classified as CSS ($\nu_\mathrm{p} < 0.5$ GHz).
For several CSS sources only upper limits on the peak frequency have been found in the literature.
The radio luminosity ($\nu L_{\nu \textrm{=5\,GHz}}$) of the sources contained in our sample varies by more than 8 orders of magnitude ($\nu L_{\nu \textrm{= 5\,GHz}}  \sim 10^{38} - 10^{46}$ erg s$^{-1}$).


\subsection{Analysis Description}
\label{sec:analysis_single_source}
We performed a dedicated analysis of each source in our sample using more than 11 years of \textit{Fermi}-LAT data between August 5, 2008, and November 1, 2019. We selected LAT data from the P8R3 source class events \citep{2018arXiv181011394B}, and P8R3\_SOURCE\_V2 instrument response functions (IRFs), in the energy range between 100\,MeV and 1\,TeV, in a region of interest (ROI) of 20$^{\circ}$ radius centered on the source position. 
The lower limit for the energy threshold is driven by the large uncertainties in the arrival directions of the photons below 100 MeV, which may be confused with the Galactic diffuse component. See \citet{2017AIPC.1792g0016P, 2018A&A...618A..22P} for a different analysis technique to solve this and other issues at low energies with \textit{Fermi}-LAT.

The analysis procedure applied in this work is mainly based on two steps. First, we investigate the $\gamma$-ray data of each individual source with a standard likelihood analysis \citep[see e.g.][]{2020A&A...635A.185P,2021ApJ...911L..11E}. The likelihood analysis, which consists of model optimization, source localization, spectrum and variability study, was performed with \texttt{fermipy}\footnote{http://fermipy.readthedocs.io/en/latest/} \citep{2017arXiv170709551W}.  Subsequently we performed a stacking analysis of the sources which were not significantly detected in the individual study, in order to investigate the general properties of the population of young radio galaxies and quasars.


\section{\textit{Fermi}-LAT results}
\label{sec:results}

\begin{figure*}[t]
\centerline{\includegraphics[width=12cm]{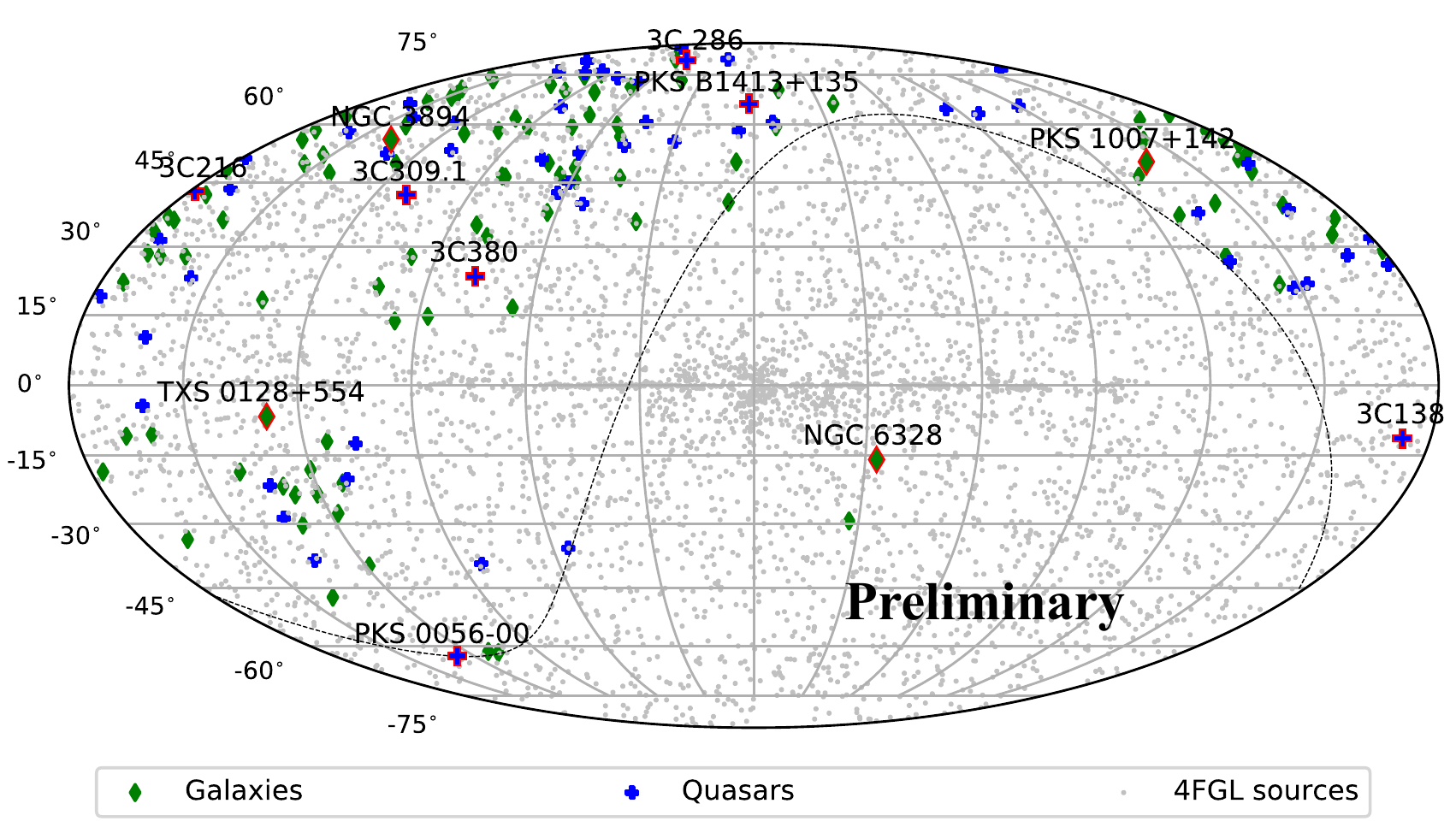}}
\caption{
Sky map, in Galactic coordinates and Mollweide projection, showing the young radio sources in our sample. The detected sources are labelled in the plot. All the 4FGL sources \citep{2020ApJS..247...33A} are also plotted, with grey points, for comparison.  \label{fig:sky_map}}
\end{figure*}

In our study we detect significant $\gamma$-ray emission (Test Statistic TS$>$ 25, corresponding to a significance $< 4.6\sigma$) at the positions of 11 young radio sources (see Fig. \ref{fig:sky_map}).

\noindent In  particular, we detect four galaxies: NGC\,6328 (LS=2 pc), NGC\,3894 (LS=10 pc), TXS\,0128+554 (LS=12 pc) and PKS\,1007+142 (LS=3.3 kpc);  and seven quasars: 3C\,138 (LS=5.9 kpc), 3C\,216 (LS=56 kpc), 3C\,286 (LS=25 kpc), 3C\,309.1 (LS=17 kpc), 3C\,380 (LS=11 kpc), PKS\,0056-00 (LS=15 kpc) and PKS\,B1413+135 (LS=0.03 kpc). 
Nine out of the 11 detected sources were present in previous \textit{Fermi}-LAT catalogs, while PKS\,0056-00 has recently been reported in the latest release of LAT sources 4FGL-DR2\footnote{https://fermi.gsfc.nasa.gov/ssc/data/access/
lat/10yr\_catalog/} \citep{2020ApJS..247...33A}.
In addition to the sources already included in the 4FGL-DR2, we report here the discovery of $\gamma$-ray emission from the young radio galaxy PKS\,1007+142 ($z$ = 0.213). PKS\,1007$+$142 presents a soft $\gamma$-ray spectrum with best-fit photon index $\Gamma = 2.55 \pm 0.18$ and flux $F =(4.65 \pm 1.55) \times 10^{-9}$ ph cm$^{-2}$ s$^{-1}$.

Considering the detected sources, all the galaxies are GPS ($\nu_\mathrm{p}$ > 0.5 GHz), while all the quasars are CSS, with the exception of the peculiar GPS PKS\,B1413$+$135.
Five quasars (3C\,138, 3C\,216, 3C\,309.1, 3C\,380 and PKS\,B1413$+$135) present significant variability (TS$_\mathrm{var}$ > 23) on a yearly scale of the $\gamma$-ray emission. The presence of $\gamma$-ray flares in these sources suggests that their high-energy emission is due to a relativistic jet and beaming effect, confirming their non-misaligned nature.

\begin{figure*}
\begin{center}
\hspace*{-0.5cm}
\rotatebox{0}{\resizebox{!}{65mm}{\includegraphics{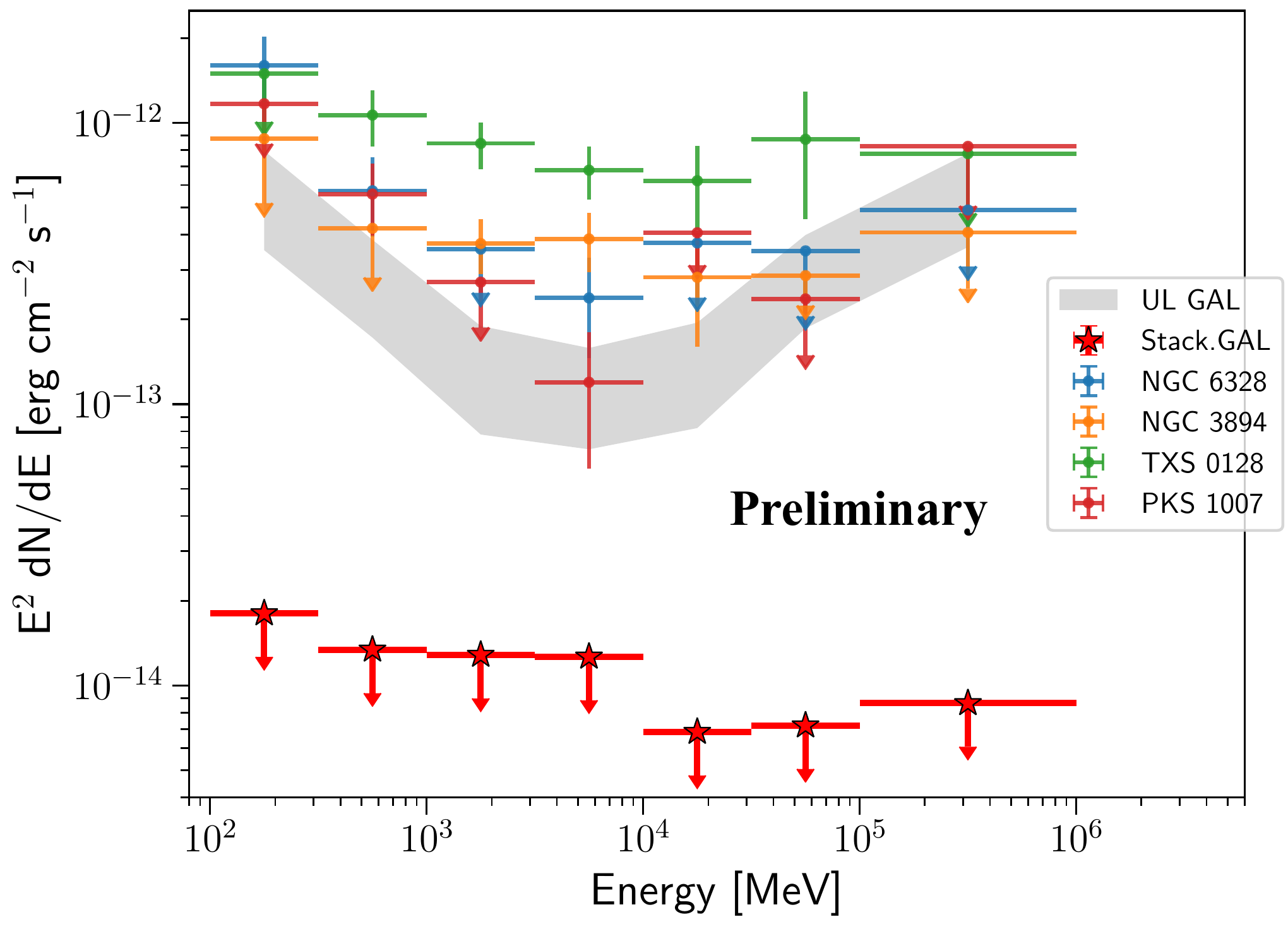}}}
\hspace{0.1cm}
\rotatebox{0}{\resizebox{!}{65mm}{\includegraphics{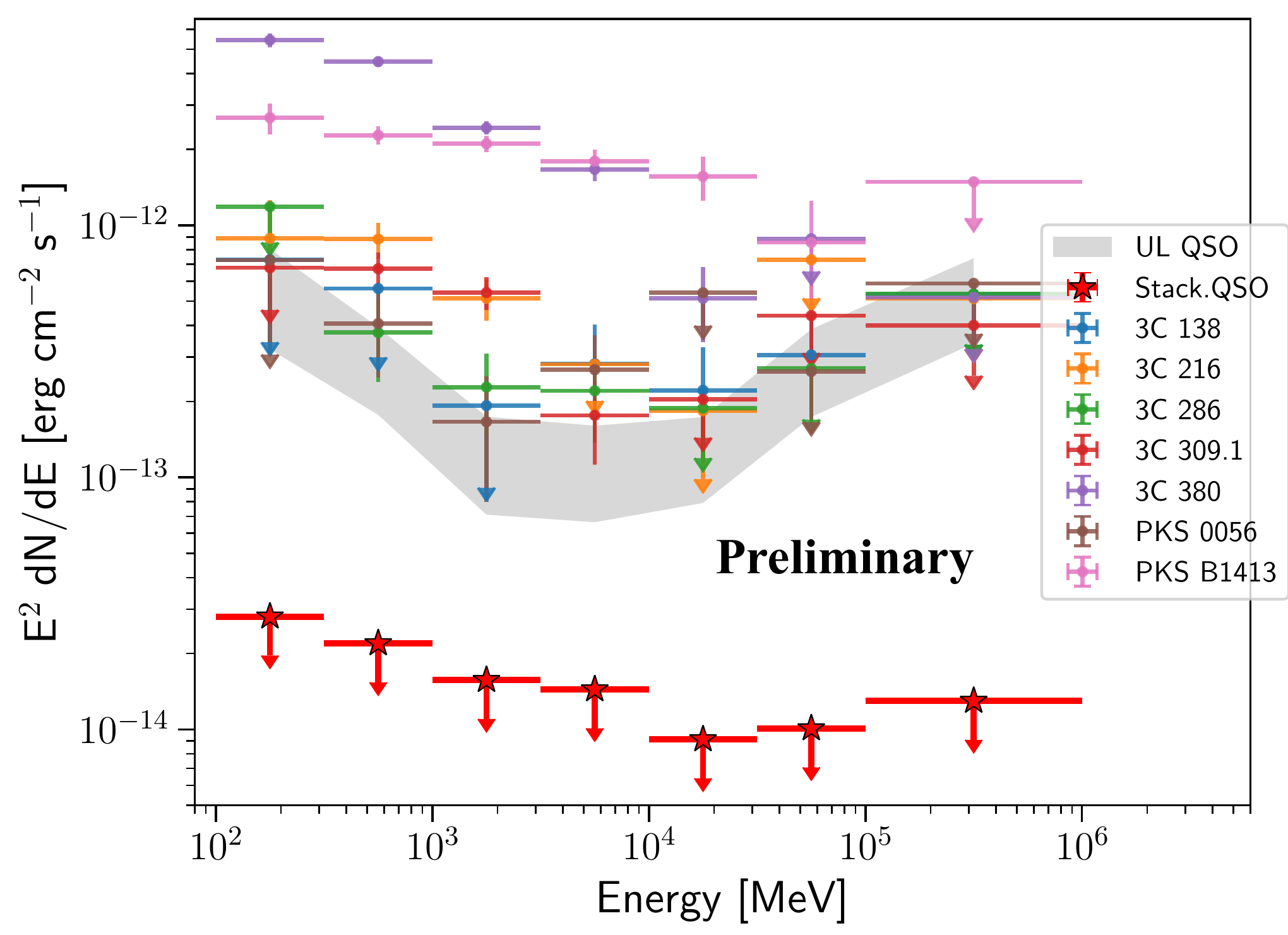}}}
\caption{  \label{fig_stacking_sed_gal} {\rm Left panel}: Upper limits for the undetected young radio galaxies as determined from the stacking analysis, compared to the averaged upper limits of the individual undetected radio galaxies (grey band) and the SEDs of the detected ones. {\rm Right panel} panel: the same as for the {\rm left} panel but for quasars.}
\end{center}
\end{figure*}

We searched for a signal from the population of the 151 undetected young radio sources by applying the stacking procedure \citep[see][ for more information on the stacking analysis]{2021MNRAS.507.4564P}.
The stacking analysis on the LAT data of the undetected sources did not result in the detection of significant emission. 
The upper limits obtained with this procedure are, however, about one order of magnitude below than those derived from the individual sources (see Fig. \ref{fig_stacking_sed_gal}).

\noindent This allowed a comparison with the model proposed by \citet{2008ApJ...680..911S}, based on an isotropic $\gamma$-ray emission from the compact lobes of young radio galaxies, excluding jet powers $(\gtrsim$10$^{42}$--10$^{43}$ erg s$^{-1}$) coupled with UV luminosities $>$ 10$^{45}$ erg s$^{-1}$. More information on the obtained results can be found in \citet{2021MNRAS.507.4564P}.

\section{Conclusion}
Before the launch of \textit{Fermi}-LAT, young radio sources were predicted to emerge as a new class of $\gamma$-ray emitting objects. However, after more than ten years of observations, only a handful of sources have been unambiguously detected  \citep{2015ApJ...810...14A,2016ApJ...821L..31M,2020ApJ...892..105A,2020A&A...635A.185P,2020ApJ...899..141L}, with the quasars playing a major role. 
The goal of our study was to investigate the $\gamma$-ray properties of young radio sources. To this end we analysed 11.3 years of \textit{Fermi}-LAT data for a sample of 162 sources. We analysed the $\gamma$-ray data of each source individually to search for a significant detection. We report the detection of 11 young radio sources, including the discovery of significant $\gamma$-ray emission from the compact radio galaxy PKS\,1007+142. In addition, we perform the first stacking analysis of this class of sources in order to investigate the $\gamma$-ray emission of the young radio sources that are undetected at high energies, without finding significant emission. The upper limits obtained with this procedure are, however, tighter than those derived from the individual sources. This enabled a comparison with the model proposed by \citet{2008ApJ...680..911S}, predicting isotropic $\gamma$-ray emission from the compact lobes of young radio galaxies. As a result we can rule out jet powers $\gtrsim$10$^{42}$--10$^{43}$ erg s$^{-1}$ coupled with UV luminosities $>$ 10$^{45}$ erg s$^{-1}$.
More information on this project can be found in \citet{2021MNRAS.507.4564P}. 

\subsection*{ACKNOWLEDGMENTS}
The \textit{Fermi}-LAT Collaboration acknowledges support from NASA and DOE (United States), CEA/Irfu, IN2P3/CNRS, and CNES (France), ASI, INFN, and INAF (Italy), MEXT, KEK, and JAXA (Japan), and the K.A. Wallenberg Foundation, the Swedish Research Council, and the National Space Board (Sweden).

\bibliography{Wiley-ASNA}

\section*{Author Biography}

\begin{biography}{
}
{\textbf{Giacomo Principe.} Graduated in Physichs in 2015 at Universit\`{a} di Padova, Italy. PhD degree in Astroparticle physichs between Jan. 2016 and Sept. 2018 at University of Erlangen-Nuremberg, graduated in March 2019. Research fellow at INAF-IRA Bologna between Oct. 2018 and Sept. 2020. Since Oct. 2020 he is a research fellow at Universit\`{a} di Trieste, Italy.}
\end{biography}

\end{document}